\documentclass[openacc]{rstransa}

\jname{rsta}
\Journal{Phil. Trans. R. Soc}

\usepackage{anyfontsize}
\usepackage[utf8]{inputenc}

\newcommand{\order}{\ensuremath{\mathcal{O}}}
\newcommand{\urlref}[2] {\href{#1}{#2}\footnote{\url{#1}, retrieved \today.}}

\begin{document}

\title{Climbing down Charney's ladder: Machine Learning and the post-Dennard era of computational climate science}

\author{
V. Balaji$^{1}$}

\address{$^{1}${Princeton University and NOAA/Geophysical Fluid
    Dynamics Laboratory, NJ, USA \\
    Institute Pierre-Simon Laplace, Paris, France}}

\subject{climate science, high performance computing, machine learning}


\corres{V. Balaji\\
\email{balaji@princeton.edu}}

\begin{abstract}
  The advent of digital computing in the 1950s sparked a revolution in
  the science of weather and climate. Meteorology, long based on
  extrapolating patterns in space and time, gave way to computational
  methods in a decade of advances in numerical weather forecasting.
  Those same methods also gave rise to computational climate science,
  studying the behaviour of those same numerical equations over
  intervals much longer than weather events, and changes in external
  boundary conditions. Several subsequent decades of exponential
  growth in computational power have brought us to the present day,
  where models ever grow in resolution and complexity, capable of
  mastery of many small-scale phenomena with global repercussions, and
  ever more intricate feedbacks in the Earth system.

  The current juncture in computing, seven decades later, heralds an
  end to what is called Dennard scaling, the physics behind ever
  smaller computational units and ever faster arithmetic. This is
  prompting a fundamental change in our approach to the simulation of
  weather and climate, potentially as revolutionary as that wrought by
  John von Neumann in the 1950s. One approach could return us to an
  earlier era of pattern recognition and extrapolation, this time
  aided by computational power. Another approach could lead us to
  insights that continue to be expressed in mathematical equations. In
  either approach, or any synthesis of those, it is clearly no longer
  the steady march of the last few decades, continuing to add detail
  to ever more elaborate models. In this prospectus, we attempt to
  show the outlines of how this may unfold in the coming decades, a
  new harnessing of physical knowledge, computation, and data.
\end{abstract}




\maketitle


\section{Introduction}
\label{sec:intro}

\begin{figure}[!ht]
  \begin{center}
    \includegraphics[width=136mm]{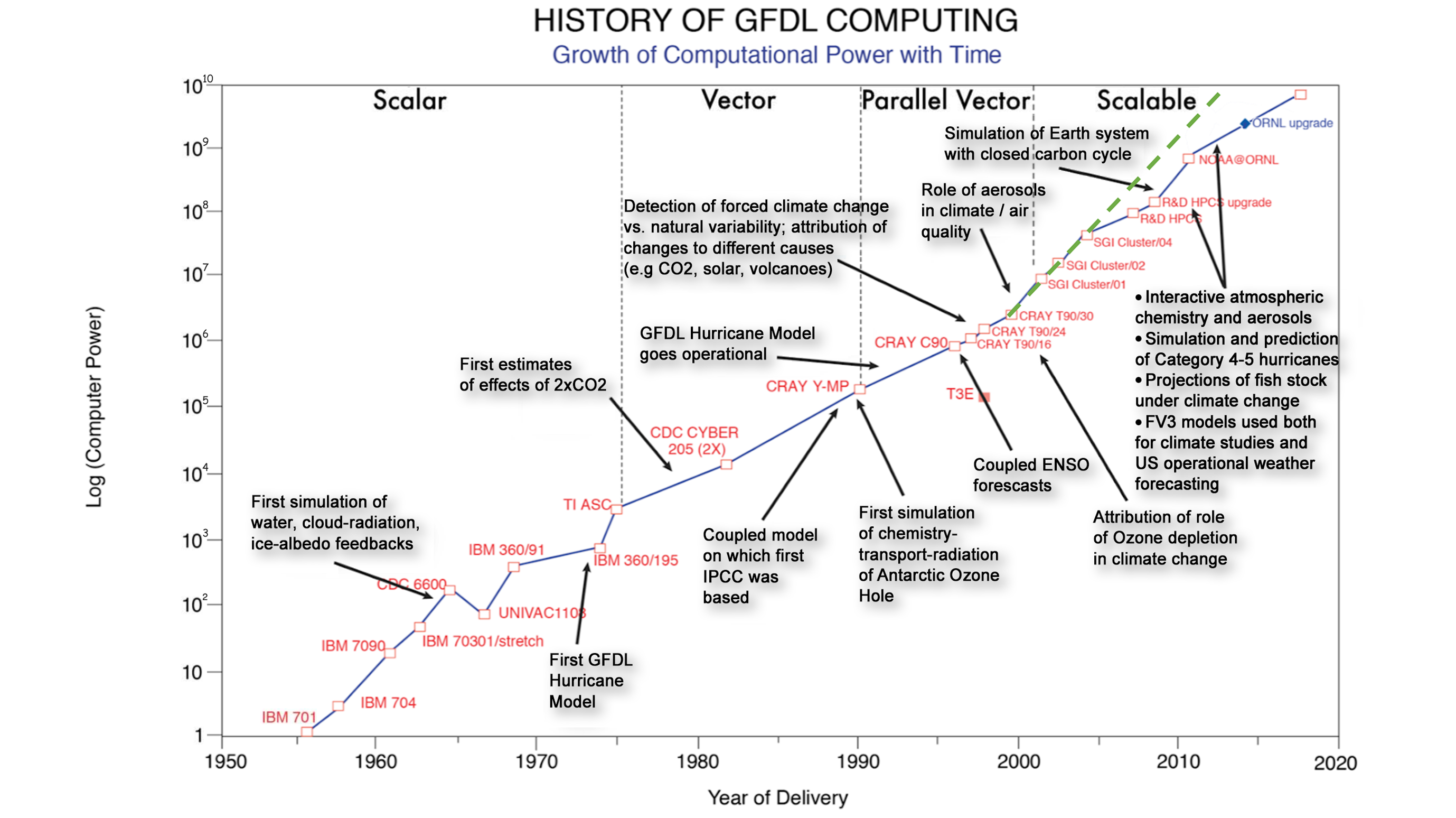}
  \end{center}
  \caption{History of computational power at the NOAA Geophysical
    Fluid Dynamics Laboratory. Computational power is measured in
    aggregate floating point operations per second, scaled so that the
    IBM 701 machine equals 1. Epochs (scalar, vector, parallel,
    scalable) show the dominant technology of the time. Landmark
    advances in climate science are shown. The green dashed line shows
    the logarithmic slope of increase in the early 2000s. Courtesy
    Youngrak Cho and Whit Anderson, NOAA/GFDL.}
  \label{fig:gfdlcomputing}
\end{figure}

The history of numerical weather prediction and climate simulation is almost exactly coincident with the history of digital computing itself \cite{ref:balaji2013d}. The story of these early days has been told many times (see e.g., \cite{ref:dahan-dalmedico2001}), including by the pioneers themselves, who worked alongside John von Neumann starting in the late 1940s. We shall revisit this history below, as some of those early debates are being reprised today, the subject of this survey. In the seven decades since those pioneering days, numerical methods have become central to meteorology and oceanography. In meteorology and numerical weather prediction, a ``quiet revolution'' \cite{ref:baueretal2015} has given us decades of steady increase in the predictive skill of weather forecasting based on models that directly integrate the equations of motion to predict the future evolution of the atmosphere, taking into account thermodynamic and radiative effects and time-evolving boundary conditions, such as the ocean and land surface. This has been made possible by systematic advances in algorithms and numerical techniques, but most of all by an extraordinary, steady, decades-long exponential expansion of computing power. Fig.~\ref{fig:gfdlcomputing} shows the expansion of computing power at one of the laboratories that in fact traces its history back to the von~Neumann era. The very first numerical weather forecast was in fact issued in 1956 from the IBM~701 \cite{ref:cressman1996} which is the first computer shown in Fig.~\ref{fig:gfdlcomputing}.

Numerical meteorology, based on a representation of real numbers in a finite number of bits, also serendipitously led to one of the most profound discoveries of the latter half of the 20th century, namely that even completely deterministic systems have limits to the predictability of the future evolution of the system \cite{ref:lorenz1963}\footnote{While this was formally known since Poincaré, Lorenz first showed it in a simple system of physical equations.}. Simply knowing the underlying physics does not translate to an ability to predict beyond a point.

Using the same dynamical systems run for very long periods of time (what von Neumann called the ``infinite forecast'' \cite{ref:smagorinsky1983}), the field of climate simulation developed over the same decades. While simple radiative arguments for CO$_2$-induced warming were advanced in the 19th century\footnote{This is generally attributed to Tyndall and Arrhenius, though earlier published research by Eunice Foote has recently come to light\cite{ref:sorenson2011}.}, numerical simulation led to a detailed understanding which goes beyond radiation and thermodynamics to include the dynamical response of the general circulation to an increase in atmospheric CO$_2$ (e.g., \cite{ref:manabewetherald1975}).

The issue of \emph{detail} in our understanding is worth spelling out.
The exorbitant increase in computing power of the last decades have
been absorbed in the adding of detail, principally in model spatial
resolution, but also in the number of physical variables and processes
simulated. That addition of detail has been shown to make significant
inroads into understanding and the ability to predict the future
evolution of the weather and climate system (henceforth Earth system).
Not only does the addition of fine spatial and temporal scales -- if
one were to resolve clouds, for example -- have repercussions at much
larger (planetary) scales, but the details of interactions between
different components of the Earth system -- ocean and biosphere for
example -- have first-order effects on the general circulation as
well.

It is the contention of this article that the continual addition of detail in our simulations is something that may have to be reconsidered, given certain physical limits on the evolution of computing hardware that we shall outline below. We may be compelled down some radically different paths.

The key feature of today's silicon-etched CMOS-based computing is that \emph{arithmetic will no longer run faster, and may even run slower, but we can perform more of it in parallel.} This has led to the revival of a computing approach which also can be traced back to the 1940s, using networks of simulated ``neurons'' to mimic processing in the human brain. The brain does in fact make use of massive parallelism but with slow ``clock speeds''. While the initial excitement around neural networks in the 1960s (e.g., the ``perceptron'' model of \cite{ref:blocketal1962}) subsided as progress stalled due to the computational limitations of the time, these methods have undergone a remarkable resurgence in many scientific fields in recent years, as the algorithms underlying learning models are ideally suited to today's hardware for arithmetic. While the meteorological community may have initially been somewhat reticent (for reasons outlined in \cite{ref:hsiehtang1998}), the last 2 or 3 years have witnessed a great efflorescence of literature applying \emph{machine learning} -- as it is now called -- in Earth system science. This special issue itself is evidence.

We argue in this article that this represents a sea change in computational Earth system science that rivals the von Neumann revolution. Indeed, some of the debates around machine learning today -- pitting ``model-free'' methods against ``interpretable AI''\footnote{AI, or artificial intelligence, is a term we shall generally avoid here in favour of terms like machine learning, which emphasize the statistical aspect, without implying insight.} for example -- recapitulate those that took place in the 1940s and 1950s, when numerical meteorology was in its infancy, as we shall show.

In subsequent sections, we revisit some of the early history of
numerical meteorology, to foreshadow some of the key debates around
machine learning today (Section~\ref{sec:hist}). In
Section~\ref{sec:dennard} we explore the issues around \emph{Dennard
  scaling} leading to the current impasse in the speed of arithmetic
\emph{in silico}. In Section~\ref{sec:ml} we look at the state of the
art of computational climate science, to see the limits of
conventional approaches, and ways forward using learning algorithms.
Finally, in Section~\ref{sec:prospects} we look at the prospects and
pitfalls in our current approaches, and outline a programme of
research that may -- or may not, as the jury is still out --
revolutionise the field by judiciously combining physical insight,
statistical learning, and the harnessing of randomness.

\section{From patterns to physics: the von Neumann revolution}
\label{sec:hist}

The development of dynamical meteorology as a science in the 20th century has been told by both practitioners (e.g., \cite{ref:lorenz1967}, \cite{ref:held2019}) and historians (e.g., \cite{ref:edwards2010}, \cite{ref:nebeker1995}). The pioneering papers of Vilhelm Bjerknes (e.g \cite{ref:bjerknes1921}) are often used as a signpost marking the beginning of dynamical meteorology, and we choose to use Vilhelm Bjerknes to highlight a fundamental dialectic that has enlivened the field since the beginning, and to this day, as we shall see below. Bjerknes pioneered the use of partial differential equations (the first use of the ``primitive equations'') to represent the state of the circulation and its time evolution, but closed-form solutions were hard to come by. Numerical methods were also immature, and basic facts about the computational stability of the methods were yet unknown. The failed attempts by Richardson \cite{ref:richardson1922} involving thousands of human ``computers'' are also well documented. One could consider it the first attempt at parallel processing, and at 64,000 ``cores'', it would be considered still quite cutting-edge today!. Bjerknes attempted a ``graphical calculus'' using drawing tools but they were imprecise at 2-3 decimal digits of precision, as stated in \cite{ref:edwards2010}, p.~87. We note this as we shall return to the issue of numerical precision later in Section~\ref{sec:dennard}. Finally abandoning the equations-based approach (global data becoming unavailable during war and its aftermath also played a role), Bjerknes reverted to making maps of air masses and their boundaries \cite{ref:bjerknes1921}. Forecasting was often based on a vast library of paper maps to find a map that resembled the present, and looking for the following sequence, what we would today recognize as Lorenz's analogue method \cite{ref:lorenz1969}. Nebeker has commented on the irony that Bjerknes, who laid the foundations of theoretical meteorology, was also the one who developed practical forecasting tools ``that were neither algorithmic nor based on the laws of physics'' \cite{ref:nebeker1995}.

We see here, in the sole person of Bjerknes, several voices in a conversation that continues to this day. One conceives of meteorology as a science, where everything can be derived from the first principles of classical fluid mechanics. A second approach is oriented specifically toward the goal of predicting the future evolution of the system (weather forecasts) and success is measured by forecast skill, by any means necessary. This could for instance be by creating approximate analogues to the current state of the circulation and relying on similar past trajectories to make an educated guess of future weather. One can have understanding of the system without the ability to predict; one can have skilful predictions innocent of any understanding. One can have a library of training data, and learn the trajectory of the system from that, at least in some approximate or probabilistic sense. If no analogue exists in the training data, no prediction is possible.

\begin{figure}[!ht]
  \begin{center}
    \includegraphics[width=136mm]{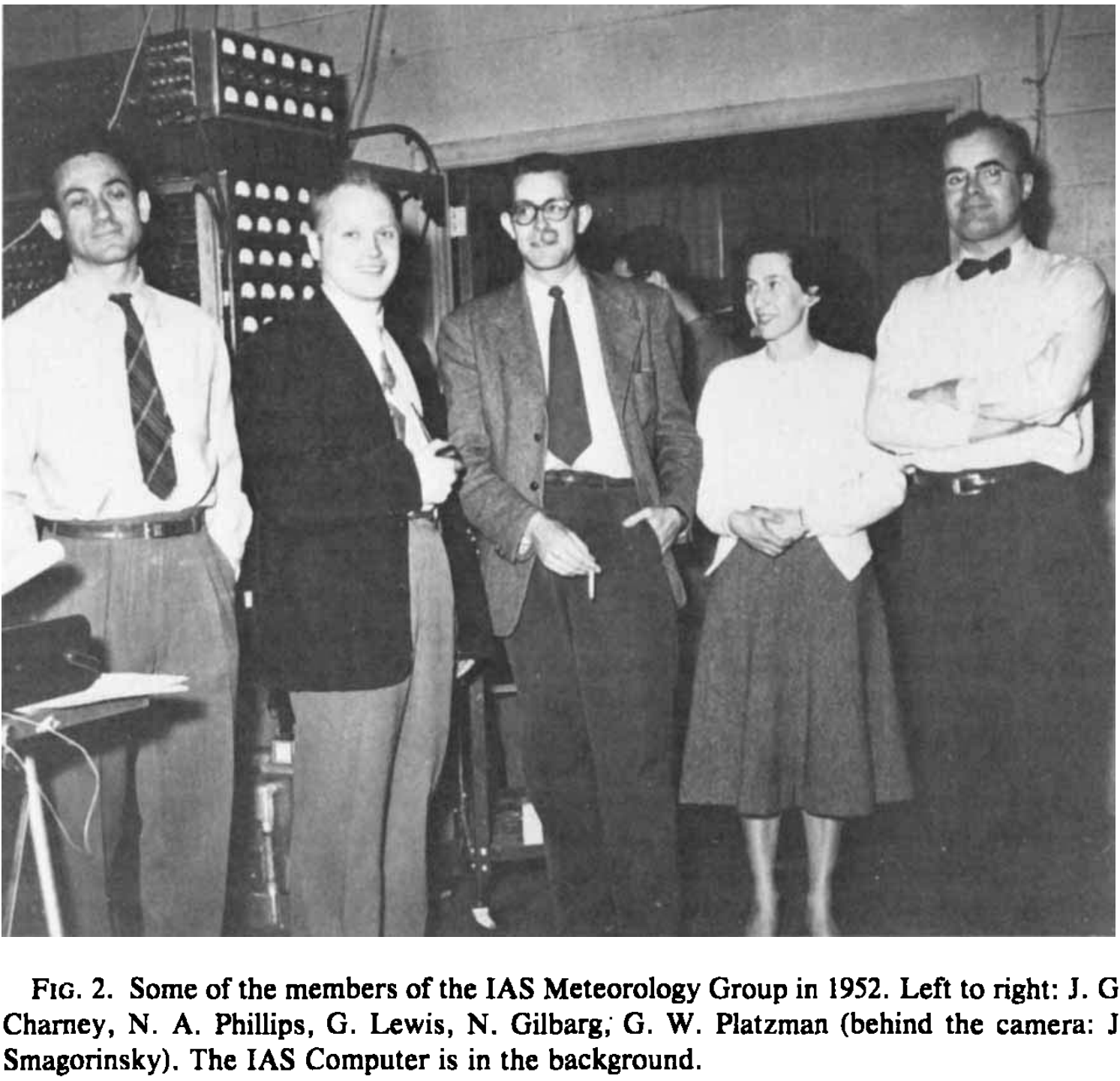}
  \end{center}
  \caption{Some of the pioneers of computational Earth system science,
    photographed by Joe Smagorinsky. From \cite{ref:smagorinsky1983}.}
  \label{fig:smagias}
\end{figure}

The physics-based approach came again to the forefront after the development of digital computing starting in the late 1940s, mainly centred around John von Neumann, Jule Charney, Joseph Smagorinsky and others at the Institute for Advanced Study (IAS) in Princeton. Once again, this story has been vividly told both by historians (\cite{ref:edwards2010}, \cite{ref:dahan-dalmedico2001}) and the participants (\cite{ref:platzman1979},\cite{ref:smagorinsky1983}), and we would not dare to try to tell it better here. Some of the pioneers are shown in this photograph taken by Smagorinsky himself (Fig.~\ref{fig:smagias}).

After Bjerknes' turn from physics to practical forecasting, that had
become the coin of the realm, and weather forecasting was entirely
based on what we would today called pattern recognition, performed by
human meteorologists. What were later called ``subjective'' forecasts
depended a lot on the experience and recall of the meteorologist, who
was generally not well-versed in theoretical meteorology, as Phillips
remarks in \cite{ref:phillips1990}. Rapidly evolving events without
obvious precursors in the data were often missed.

Charney's introduction of a numerical solution to the barotropic vorticity equation \cite{ref:charneyetal1950}, and its execution on ENIAC, the first operational digital computer, essentially led to a complete reversal of fortune in the race between physics and pattern recognition. Programmable computers (where instructions were loaded in as well as data) came soon after, and the next landmark calculation of Phillips \cite{ref:phillips1956} came soon thereafter. Charney had spoken of ``climbing the ladder'' of a hierarchy of models of increasing complexity, and the concept of the model hierarchy is something we shall revisit later as well. The Phillips 2-layer model was the next rung on the ladder.

It was not long before forecasts based on simplified numerical models outperformed subjective forecasts. Forecasting skill, measured by errors in the 500-hPa geopotential height, was clearly better in the numerical forecasts after Phillips' breakthrough (see Figure~1 in \cite{ref:shuman1989}). Forecasts based on integrating forward prognostic equations are now so taken for granted that Edwards remarks in \cite{ref:edwards2010} that he found it hard to convince some of the scientists he met in the 1990s that it was decades since the founding of theoretical meteorology before physics outdid simple heuristics and theory-free pattern recognition in forecast skill.

The same tools, numerical models on a digital computer, were quickly also put to use to see what long term fluctuations of weather might look like, von Neumann's ``infinite forecast''. Models of the ocean circulation had begun to appear (e.g \cite{ref:munk1950}) showing low-frequency (by atmospheric weather standards) variations, and the significance of atmosphere-ocean coupling had been observed \cite{ref:namias1959}. The first coupled model of Manabe and Bryan \cite{ref:manabebryan1969} appeared in Smagorinsky's lab. Smagorinsky had remarked that while the ``deterministic nonperiodic'' behaviour \cite{ref:lorenz1963} of geophysical fluids, which we now know as chaos, placed limits on predictability, the statistics of weather fluctuations in the asymptotic limit could still be usefully studied \cite{ref:smagorinsky1983}, and his laboratory soon became a centrepiece of this body of research. Within a decade, such models were the basic tools of the trade for studying the asymptotic equilibrium response of Earth system to changes in external forcing, the new field of computational climate science.

The practical outcomes of these studies, namely the response of the climate to anthropogenic CO$_2$ emissions, raised public alarm with the publication of the Charney Report in 1979 \cite{ref:charneyetal1979}. Around the same time, a revolution in technology made computing cheap and ubiquitous, a mass market commodity. This is seen in Fig.~\ref{fig:gfdlcomputing} as a transition from specialised ``vector'' processors to parallel clusters built of ``commodity'' components. Computational climate science, now with planetary-scale societal ramifications, became a rapidly growing field expanding across many countries and laboratories, which could all now aspire to the scale of computing required to work out the implications of anthropogenic climate change. There was never enough computing: it was clear for example that clouds were a major unknown in the system (as noted already in the Charney Report) and were (and still are, see \cite{ref:schneideretal2017}) well below the resolution of models able to exploit the largest computers available. The models were resource-intensive, ready to soak up every last cycle of computation and byte of memory. A more sophisticated understanding of the Earth system also began to bring more processes into the simulations, now an integrated whole with physics, chemistry and biology components. We were still climbing Charney's hierarchy and adding complexity, but quite often the new components, were empirical, sometimes well-founded in observations in nature or in the laboratory, but often curve-fits to observations that were usually inadequate in time and space. Ecosystems and the turbulent planetary boundary layer are prime examples.

While the hierarchy ladder in the early days could be described theoretically in terms of the non-dimensional numbers that governed which terms to include or neglect in the equations, a standard approach in fluid dynamics, newer rungs in the ladder were governed by dimensional numbers in equations whose structure itself was empirically determined, with parameters poorly constrained by data, or indeed with no observable counterpart in nature. When these components are assembled into a coupled system, we are left with many errors and biases that are minimised during a further ``tuning'' or calibration phase \cite{ref:hourdinetal2017} varying some of these free parameters. It has been shown for example that a model's skill at reproducing the 20th century temperature record can be tuned up or down without violating fidelity to the process-level empirical constraints \cite{ref:golazetal2013}. Some obvious ``fudge factors'' such as flux adjustments \cite{ref:shackleyetal1999} have been eliminated over the years. However, recalcitrant errors remain. The so-called ``double-ITCZ'' bias for example, which, despite glimmers of progress (see varied explanations in \cite{ref:xiangetal2017}, \cite{ref:wittenbergetal2018}, \cite{ref:woelfleetal2019}) has remained stubbornly resistant to any amount of reformulation or tuning across many generations of climate models (\cite{ref:lin2007},\cite{ref:lixie2014},\cite{ref:tiandong2020}). It is contended by many that no amount of fiddling with parameterizations can correct some of these long standing biases, and only direct simulation resolving key features is likely to lead to progress (e.g \cite{ref:palmerstevens2019}, Box~2).

The revolution begun by von Neumann and Charney at the IAS, and the
subsequent decades of exponential growth in computing shown in
Fig.~\ref{fig:gfdlcomputing}, have led to tremendous leaps forward as
well as more ambiguous indicators of progress. What initially looked
like a clear triumph of physics allied with computing and algorithmic
advances now shows signs of stalling, as the accumulation of
\emph{detail} in the models -- both in resolution and complexity --
leads to some difficulty in the interpretation and mastery of model
behaviour. Exponential growth curves in the real world eventually
turn sigmoid, and this may be true of Earth system modelling as well.
More concretely, the exponential growth of
Fig.~\ref{fig:gfdlcomputing} is also levelling off, with no immediate
recourse, as we show below in Section~\ref{sec:dennard}. This is
leading to a turn in computational climate science that may be no less
far-reaching than the one wrought in Princeton in 1950.

\section{The end of Dennard scaling: slower arithmetic, but more of it}
\label{sec:dennard}

The state of play in climate computing at the cusp of the machine
learning explosion was reviewed in \cite{ref:balaji2015}. Earth system
modeling faces certain intrinsic problems which keeps it from
realizing even the potential of current computing. Considerable
creative energy is being devoted to this problem, as outlined in
\cite{ref:balaji2015}, but we shall not revisit those efforts here.
Suffice it to say that these issues are not simply a matter of better
software engineering, but fundamental aspects of the algorithms of
fluid dynamics. But for now we restrict ourselves to the physical
limits of current computing technology, and how we may adapt to life
on the cusp of this technological transition..

Over almost four decades of the commodity microprocessor revolution alluded to earlier, there had been an extraordinary and relentless expansion in computing capacity, as seen in Fig.~\ref{fig:gfdlcomputing}. This is quite often referred to as ``Moore's Law'', based on Gordon Moore's observation that the number of transistors, the basic building blocks of digital computers, in a given area of silicon substrate doubles every 18~months, as we make miniaturization gains in Complementary Metal-Oxide-Silicon (CMOS) fabrication. Underlying Moore's Law is the physics of \emph{Dennard scaling} \cite{ref:dennardetal1974}. The microprocessor revolution is fueled by our ability to etch circuits at finer scale with each cycle in fabrication, resulting in faster switching (and thus the speed of an arithmetic operation) without any increase in power requirements, as shown in the last row of Table~\ref{tab:dennard}.

\begin{table}[!ht]
  \begin{center}
    \begin{tabular}[c]{||c|c||} \hline\hline
      Device Parameter & Scaling Factor \\ \hline
      Doping concentration & $\kappa$ \\
      Transistor Dimension & 1/$\kappa$ \\
      Voltage $V$ & 1/$\kappa$ \\
      Current $I$ & 1/$\kappa$ \\
      Capacitance $C$ & 1/$\kappa$ \\
      Delay time per circuit $VC/I$ & 1/$\kappa$ \\
      Power dissipation per circuit $VI$ & 1/$\kappa^2$ \\
      Power density $VI/A$ & 1 \\ \hline\hline
    \end{tabular}
  \end{center}
  \caption{The basis for Dennard scaling. With every shrink cycle in
    transistor fabrication, we gain the linear shrink factor $\kappa$
    in circuit switching speed, while maintaining power density
    constant. Adapted from Table~1 in \cite{ref:dennardetal1974}.}
  \label{tab:dennard}
\end{table}

As transistor dimensions continue to shrink, Dennard scaling is
approaching its physical limits, as noted by \cite{ref:markov2014}:
for example at the current 5~nm fabrication dimension, transistors are
about 30~atoms across. At the physical limit of CMOS, Dennard scaling
breaks down \cite{ref:chienkaramcheti2013}. First of all, clock speeds
no longer increase. More critically, power density no longer stays
constant, decreasing performance per watt. The increase in power
dissipation per unit area also implies increased heat dissipation,
leading to the phenomenon of ``dark silicon''
\cite{ref:esmaeilzadehetal2013}, where large sections, often more than
50\% of the chip surface have to be turned off (no current) at any
given moment, in order to stay within safe thermal limits. This means
that the number of operations per cycle is actually well below what is
theoretically possible.

\begin{figure}[!ht]
  \begin{center}
    \includegraphics[width=136mm]{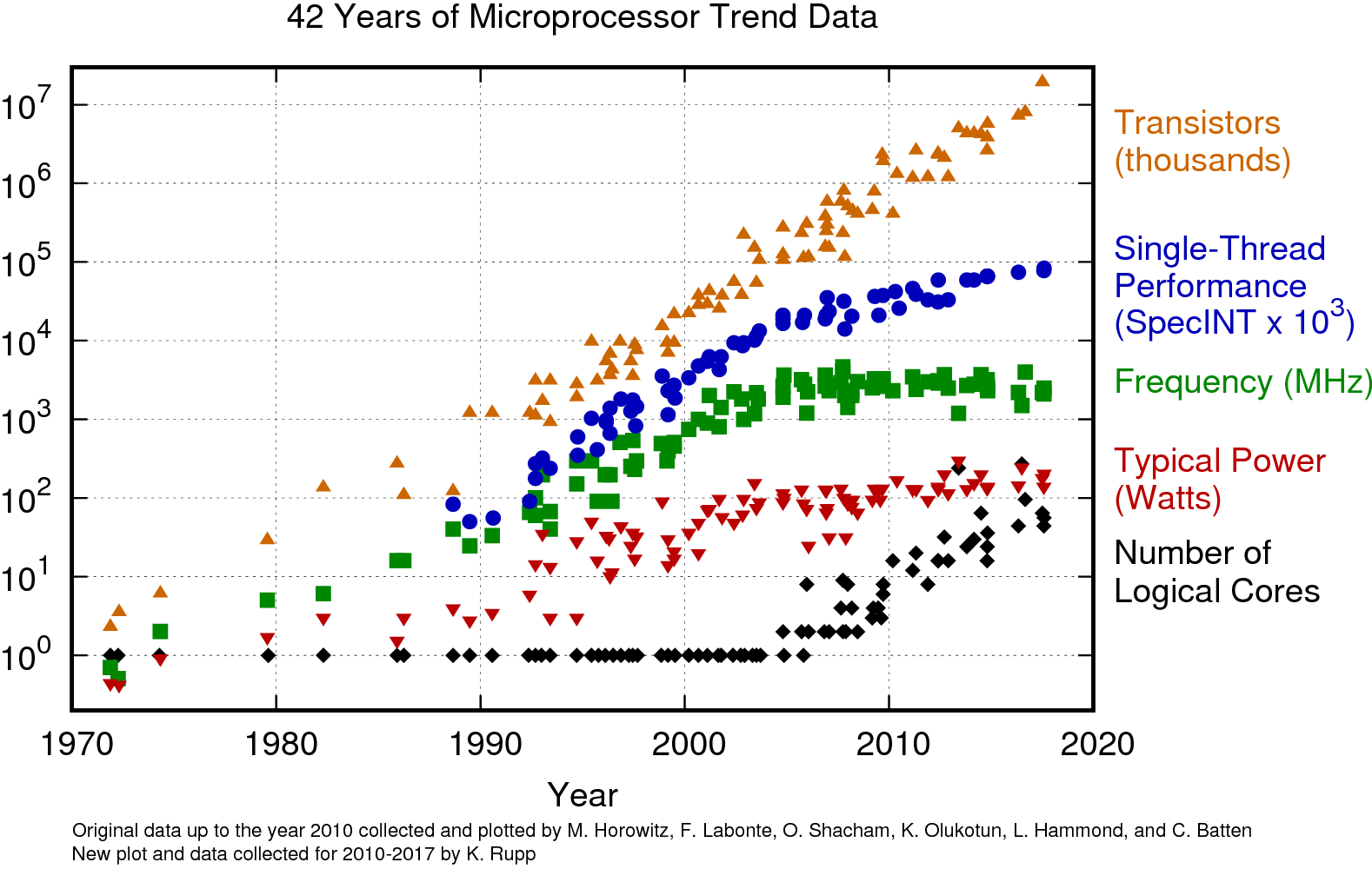}
  \end{center}
  \caption{Dennard scaling tails off at the end of four decades of
    microprocessor miniaturization. From
    \href{https://www.karlrupp.net/2018/02/42-years-of-microprocessor-trend-data/}{42
      Years of Microprocessor Trend Data}, courtesy Karl Rupp.}
  \label{fig:dennard}
\end{figure}

The sigmoid taper of exponential growth in microprocessor speed is clearly seen in Fig.~\ref{fig:dennard}. While the number of transistors still remains on a doubling trajectory, the actual arithmetic speed (chip frequency in MHz) and wattage have levelled off. The transistors now go to increase the number of logical cores on the chip. As a result, arithmetic speed is now static or may even become slower to maintain the power and cooling envelope, but more arithmetic can be concurrently executed. The difficulties of continuing to increase resolution under these constraints have been described in \cite{ref:scharetal2020}.

So, certain physical limits of current computing have been reached.
While there are efforts to imagine radically novel computing methods,
including ``non-von Neumann'' methods such as quantum or neuromorphic
computing \cite{ref:conteetal2017}, those are still on the drawing
board. Interesting forays edging toward ``inexact computing''
(\cite{ref:palem2014},\cite{ref:palmer2015}) remain in exploratory
stages as well.

It is clear however that absent some unforeseen advance, current approaches to Earth system modeling will not be advancing in the same fashion as the last several decades. An analysis of the potential performance of a global cloud-resolving model on an exascale machine in \cite{ref:neumannetal2019} showed that it would potentially take an entire such machine to run that model at a speed of 1 simulated year per day (SYPD). Climate science requires not just computing \emph{capability} (SYPD) but also \emph{capacity}, the ability to run multiple copies of the model to sample various kinds of uncertainty. A suite of typical climate experiments aimed at studying the response of the model to various perturbations is measured in \order(10,000) years. Simply developing and testing a model, including the calibration described in Section~\ref{sec:hist}, may cost 5 times as much \cite{ref:adcroftetal2019}. Similar costs obtain in establishing a model's predictive skill by running a suite of retrospective hindcasts. Furthermore, the model studied in \cite{ref:neumannetal2019} has vastly less \emph{complexity} (defined in \cite{ref:balajietal2017} as the number of distinct physical variables simulated by a model) than a typical workhorse model. It is clear that the additional of detail, in resolution and complexity, to models cannot continue as before. A fundamental rethinking of the decades-long climb up the Charney ladder is long overdue.

Just around the time the state of play in climate computing was reviewed in \cite{ref:balaji2015}, the contours of the revival of machine learning (ML) using artificial neural networks (ANNs) were beginning to take shape. Deep learning (DL) using multiple neuronal layers were showing significant skill in many domains. As noted in Section~\ref{sec:intro}, ANNs existed alongside the physics-based models of von Neumann and Charney for decades, but may have languished as the computing power and parallelism were not available. The new processors emerging at the right of Fig.~\ref{fig:dennard} in the twilight of Dennard scaling, are ideally suited to ML: the typical DL computation consists of dense linear algebra, scalable almost at will, able to reduce memory bandwidth at reduced precision without loss of performance. Processors such as the Graphical Processing Unit (GPU) and most especially the TPU (tensor processing unit) showed themselves capable of running a typical DL workload at close to the theoretical maximum performance of the chip \cite{ref:jouppietal2017}. There are many challenges to executing conventional equation-based arithmetic on these chips, not least of which is their low precision, often as low as 3 digits, the same as for the manual arithmetic that limited Bjerknes, see Section~\ref{sec:hist}. While continuing to explore low-precision arithmetic (e.g \cite{ref:chantryetal2018}), we have begun to explore ML itself in the arsenal of Earth system modeling. We turn now in Section~\ref{sec:ml} to an assess the potential of ML to show us a way out of the current computing impasse.

\section{Learning physics from data}
\label{sec:ml}

The articles in this special issue form a wide spectrum representing
the state of the art in the use of ML in Earth system science, and we
do not propose to offer a broad or comprehensive review. Instead, we
aim to demonstrate using a judicious selection of a few threads of
research from the current literature how Earth system modeling's turn
toward ML reprises some of the fundamental issues that arose in the
pioneering era, outlined above in Section~\ref{sec:hist}.

Recall V. Bjerknes's turn away from theoretical meteorology upon finding that the tools at his disposal were not adequate to the task of making predictions from theory. It is possible that the computational predicament we now find ourselves in, as outlined above in Section~\ref{sec:dennard}, is a historical parallel, and we too shall turn toward practical, theory-free predictions. An example would be the prediction of precipitation from a sequence of radar images \cite{ref:agrawaletal2019}, where ``optical flow'' (essentially, extrapolating various optical transformations such as translation, rotation, stretching, intensification) is compared and found competitive with persistence and short term model-based forecasts. Similarly,

ML methods have shown exceptional forecast skill at longer timescales,
including breaking through the ``spring barrier'' (the name given to a
dramatic reduction in forecast skill in models initialised prior to
boreal spring) in ENSO predictability \cite{ref:hametal2019}.
Interestingly, the Lorenz analogue method \cite{ref:lorenz1969} also
shows longer term ENSO skill (with no spring barrier) compared to
dynamical models \cite{ref:dingetal2019}, a throwback to the early
days of forecasting as described in Section~\ref{sec:hist}. These and
other successes in purely ``data-driven'' forecasting (though the ENSO
papers use model output as training data) have led to speculation in
the media that ML might indeed make physics-based forecasting obsolete
(see for example,
\urlref{https://www.hpcwire.com/2020/04/27/could-machine-learning-replace-the-entire-weather-forecast-system/}{Could
  Machine Learning Replace the Entire Weather Forecast System?} in
\emph{HPCWire}). ML methods (in this case, recurrent neural networks)
have also shown themselves capable of reproducing a time series from
canonical chaotic systems with predictability beyond what dynamical
systems theory would suggest, e.g., \cite{ref:pathaketal2018} (which
indeed explicitly makes a claim to be ``model-free''),
\cite{ref:chattopadhyayetal2020}. Does this mean we have come full
circle on the von Neumann revolution, and return to forecasting from
pattern recognition rather than physics? The answer of course is
contingent on the presumption that the training data in fact is
comprehensive and samples all possible states of the system. For the
Earth system, this is a dubious proposition, as there is variability
on all time scales, including those longer than the observational
record itself. A key issue for all data-driven approaches is that of
\emph{generalisability} beyond the confines of the training data.

Turning to climate from weather, we look at the aspects of Earth system models that while broadly based on theory, are structured around an empirical formulation rather than from first principles. These are areas obviously ripe for a more directly data-driven approach. These often are based around the parameterised components of the model that deal with ``sub-gridscale'' processes operating below the truncation imposed by discretisation. A key feature of geophysical fluid flow is the 3-dimensional turbulence cascade continuous from planetary scale down to the Kolmogorov length scale (see e.g., \cite{ref:nastromgage1985}, Fig.~1), which must be truncated somewhere for a discrete numerical representation. ML-based representation of subgrid turbulence is one area receiving considerable attention now \cite{ref:duraisamyetal2019}. Other sub-gridscale aspects particular to Earth system modeling where ML could play a role include radiative transfer and the representation of clouds, which now has a rich literature, including in this special issue. As noted just above, ``data-driven'' methods often rely on the output of models. This is particularly so for the case of sub-gridscale physics, where learning often involves \emph{emulating} the behaviour of models in which the phenomenon is resolved.

ANNs have the immediate advantage of often being considerably faster
than the component they replace
(\cite{ref:krasnopolskyetal2005},\cite{ref:raspetal2018}), thus
directly responsive to the computational challenge laid down in
Section~\ref{sec:dennard}. They additionally have the advantage of
being differentiable, which sub-grid physics often is not: this has a
key advantage in the use of data assimilation (DA) techniques to
constrain model trajectory. The formal equivalences between DA and ML
have been demonstrated in \cite{ref:kovachkistuart2019}, for instance
between adjoint models used in DA and the back-propagation technique
in ML. Further, the calibration procedures of
\cite{ref:hourdinetal2017} are very inefficient with full ESMs, and
may be considerably accelerated using \emph{emulators} derived by
learning \cite{ref:williamsonetal2013}.

ML nonetheless poses a number of challenging questions that we are now actively addressing. The usual problems of whether the data is representative and comprehensive, and on generalisability of the learning, continue to apply. There is a conundrum in deciding where the boundary between being physical knowledge driven and data driven lies. We outline some key questions being addressed in the current literature, and not least in this special issue. We take as a starting point, a particular model component (such as atmospheric convection or ocean turbulence) that is now being augmented by learning.

Do we take the structure of the underlying system as it is now, and
use learning as a method of reducing parametric uncertainty? Emerging
methods potentially do allow us to treat structural and parametric
error on a common footing (e.g \cite{ref:williamsonetal2015}), but we
still may choose to go either structure-free, or attempt to discover
the structure itself.

If we divest ourselves of the underlying structure of equations, we have a number of issues to address. As the learning is only as good as the training data, we may find that the resulting ANN violates some basic physics (such as conservation laws), or does not generalize well \cite{ref:boltonzanna2019}. This can be addressed by a suitable choice of basis in which to do the learning, as subsequent work \cite{ref:zannabolton2020} shows. A similar consideration was seen in \cite{ref:ogormandwyer2018}, which when trained on current climate failed to generalize to a warmer climate, but the loss of generalizability could be addressed by a suitable choice of input basis, using lapse rates rather than temperatures for example (O'Gorman, personal communication)..

Ideally, we would like to go much further and actually learn the underlying physics. There have been attempts to learn the underlying equations for well-known systems (\cite{ref:schmidtlipson2009},\cite{ref:bruntonetal2016}) and efforts underway in climate modeling as well, to learn the underlying structure of parameterizations from data. In this context, ``learning the physics'' means, for instance, starting from data and writing down closed-form equations for the effect of unresolved physics on resolved-scale tendencies. This is by its nature seeking a \emph{sparse} representation, as a closed-form equation is one one can write down, and must have a limited number of terms on the right hand side. For tests of the canonical systems in the articles cited above, this is guaranteed, as the systems we are learning are known to be sparse in this sense. The challenge is to attempt to learn it for systems where a sparse representation is not known or guaranteed to exist. \cite{ref:zannabolton2020} is an interesting step forward in this direction. However, the current methods still face some limitations. In particular the methods only seek representations within a library of possible eigenfunctions that can be used. Ingenuity and physical insight still play a role in finding the basis that yields a sparse, or parsimonious, representation.

Finally, we pose the problem of coupling. We have noted earlier the problem of calibration of models, which is done first at the component level, to bring each individual process within observational constraints, and then in a second stage of calibration against systemwide constraints, such as top of atmosphere radiative balance \cite{ref:hourdinetal2017}. The issue of the stability of ANNs when integrated into a coupled system is also under active study at the moment (e.g \cite{ref:brenowitzetal2020}). The question of whether tightly coupled subsystems should be jointly or separately learned still remains an open question.

\section{Climbing down the ladder: prospects for ML-based Earth system
  models}
\label{sec:prospects}

We have highlighted in this article a historical progression in Earth
system modeling, which we describe here as the addition of
\emph{detail} to our models, in the form of resolution and complexity.
While this has resulted in tremendous leaps in understanding, there
have been some aspects where we have failed to advance for decades.
In most accounts, this can be traced to failures in representation of
the form of sub-gridscale phenomena such as clouds. The solution may
come when we in fact no longer need to seek representations, as we
shall be resolving such phenomena directly. We have noted possible
setbacks to this approach given physical limits on computing technology.

We have gazed into the computational abyss before us, and seen how machine learning may offer ways to adapt to what today's machines do best, which is statistical machine learning. It is a transition in our approach to predicting the Earth system that is potentially as far-reaching as that of von Neumann and Charney, algorithms that learn rather than do what they're told. How this will unfold is yet to be seen.

At first glance, it may appear as though we are turning our backs on
the von Neumann revolution, going back from physics to simply seeking
and following patterns in data. While such black-box approaches may
indeed be used for certain activities, we are seeing many attempts to
go beyond those, and ensure that the learning algorithms do indeed
respect physical constraints even if not present in the data.

Even ``data-driven'' methods quite often rely on the output of models, as noted at various points in the text: indeed reconstructed historical observations (reanalysis datasets) are reliant on models to produce a physically consistent multivariate dataset out of diverse observations. The observed climate record is also too short for unambiguously separating natural from forced variations: we rely on models to supplement that record. It is unlikely that raw data alone will suffice to train ML algorithms. Models that operate at the limit of resolution on the largest computers available will indeed be among the tools of the trade, but they will likely not be the workhorses of modeling, which require many runs to see how they respond to perturbations. Building reduced order models, lower in resolution and complexity, may well become one of the principal uses of the extreme-scale models.

It was said of Charney's first computation that he knew what to leave
out for a feasible calculation. Since then, computing power has grown,
and the models have climbed the ladder of a hierarchy of complexity.
But as Held has remarked \cite{ref:held2005}, it is necessary to
descend the hierarchy as well, to pass from simulation to
understanding. If ML-based modeling needs a manifesto, it may be this:
to \emph{learn from data not just patterns, but simpler models},
climbing down Charney's ladder. The path down the ladder will not be
an exact reprise, but will include what we learned on the rungs on the
way up. The vision is that these models will leave out the details not
needed in an understanding of the underlying system, and learning
algorithms will find for us underlying ``slow manifolds''
\cite{ref:lorenz1992}, and the basis variables that yield a succinct,
or parsimonious, mapping. That is the challenge before us.

\enlargethispage{20pt}


\dataccess{Data and software for Fig.~\ref{fig:dennard} is available at the \urlref{https://github.com/karlrupp/microprocessor-trend-data}{Microprocessor Trend Data Repository} on Github, under a Creative Commons license. We are very grateful to Karl Rupp for graciously making this available. The specific data for this picture is available at Zenodo with DOI 10.5281/zenodo.3947824 \cite{ref:rupp2020}.}


\competing{The author(s) declare that they have no competing
  interests.}

\funding{ V. Balaji is supported by the Cooperative Institute for Modeling the Earth System, Princeton University, under Award NA18OAR4320123 from the National Oceanic and Atmospheric Administration, U.S. Department of Commerce, and by the \emph{Make Our Planet Great Again} French state aid managed by the Agence Nationale de Recherche under the ``Investissements d'avenir'' program with the reference ANR-17-MPGA-0009. }

\ack{ I thank Alistair Adcroft, Isaac Held, Nadir Jeevanjee and
  Syukuro Manabe and 2 anonymous reviewers for comments on early
  drafts of this manuscript. I have benefited from conversations and
  discussions with Alistair Adcroft, Tom Bolton, Noah Brenowitz, Chris
  Bretherton, Hannah Christensen, Fleur Couvreux, Julie Deshayes,
  Peter D\"uben, Jonathan Gregory, Isaac Held, Fr\'ed\'eric Hourdin,
  Redouane Lguensat, Tim Palmer, Tapio Schneider, Anna Sommer, Maike
  Sonnewald, Danny Williamson, Laure Zanna, and many others, although
  any errors in the interpretation of their insights are entirely
  mine. I would also like to thank the organizers of the 2019 Oxford
  Workshop on Machine Learning for Weather and Climate for the
  opportunity to participate and contribute to this special issue. }

\disclaimer{ The statements, findings, conclusions, and recommendations are those of the authors and do not necessarily reflect the views of Princeton University, the National Oceanic and Atmospheric Administration, the U.S. Department of Commerce, or the French Agence Nationale de Recherche. }

\bibliographystyle{ieeetr}
\bibliography{refs}

\end{document}